\documentclass{PoS}

\usepackage[utf8]{inputenc}
\usepackage{graphicx}
\usepackage{slashed}
\usepackage{color}
\usepackage{amsmath}
\usepackage{amssymb}

\usepackage{multirow}
\usepackage{tabularx}
\newcolumntype{C}[1]{>{\hsize=#1\hsize\centering\arraybackslash}X}%

\newcolumntype{Z}{r<{\hspace{3mm}}}
\definecolor{mygreen}{rgb}{0,0.7,0}

\newcommand{\eps}{\epsilon}
\DeclareMathOperator{\tr}{\rm tr}

\title{Applications of integrand reduction to two-loop five-point scattering amplitudes in QCD}

\ShortTitle{Two-loop five-point amplitudes}

\author{\speaker{Simon Badger}\\
        Institute for Particle Physics Phenomenology, Department of Physics, Durham University, Durham DH1 3LE, United Kingdom\\
        E-mail: \email{simon.d.badger@durham.ac.uk}}

\author{Christian Br\o nnum-Hansen\\
        Higgs Centre for Theoretical Physics, School of Physics and Astronomy, The University of Edinburgh, Edinburgh EH9 3JZ, Scotland, UK\\
        E-mail: \email{bronnum.hansen@ed.ac.uk}}

\author{Thomas Gehrmann\\
        Physik-Institut, Universit\"at Z\"urich, Winterthurerstr. 190, CH-8057 Z\"urich, Switzerland \\
        E-mail: \email{thomas.gehrmann@uzh.ch}}

\author{Heribertus Bayu Hartanto\\
        Institute for Particle Physics Phenomenology, Department of Physics, Durham University, Durham DH1 3LE, United Kingdom\\
        E-mail: \email{heribertus.b.hartanto@durham.ac.uk}}

\author{Johannes Henn\\
        PRISMA Cluster of Excellence, Johannes Gutenberg University, 55128 Mainz, Germany\\
        MPI f\"ur Physik, Werner-Heisenberg-Institut, M\"unchen, Germany \\
        Email: \email{henn@uni-mainz.de}}

\author{Nicola Adriano Lo Presti\\
        Institute for Particle Physics Phenomenology, Department of Physics, Durham University, Durham DH1 3LE, United Kingdom\\
        Email: \email{nicola.a.lo-presti@durham.ac.uk}}

\author{Tiziano Peraro\\
        PRISMA Cluster of Excellence, Johannes Gutenberg University, 55128 Mainz, Germany\\
        E-mail: \email{peraro@uni-maniz.de}}

\abstract{We review the current state-of-the-art in integrand level reduction for five-point scattering amplitudes
at two loops in QCD. We present some benchmark results for the evaluation of the leading colour two-loop five-gluon amplitudes
in the physical region as well as the partonic channels for two quarks and three gluons and four quarks and one gluon.}

\FullConference{Loops and Legs in Quantum Field Theory (LL2018)\\
		29 April 2018 - 04 May 2018\\
		St. Goar, Germany}

\begin{document}
\section{Introduction}

In these proceedings we review some recent developments in integrand reduction
methods for two-loop amplitudes and their application to five particle
amplitudes in massless QCD. NNLO QCD corrections to processes such as $pp\to
3j$, $pp\to H+2j$ and $pp\to W+2j$ are currently a high priority within the
community~\cite{Bendavid:2018nar} where many measurements are already dominated
by theoretical uncertainties. New methods to overcome the complexity of these
amplitudes have been an on-going effort for many years and useful techniques
have been proposed from many directions. High multiplicity multi-loop
computations in maximally supersymmetric theories like $\mathcal{N}=4$ super
Yang-Mills (sYM) are now common place where the current state of the art is
planar 6-point amplitudes at 5-loops~\cite{Caron-Huot:2016owq} with integrand
representations known in principle to any loop order~\cite{ArkaniHamed:2010kv}.
Understanding $\mathcal{N}=4$ sYM amplitudes has often provided valuable
information when attempting to construct efficient computational strategies for
more complicated and phenomenologically relevant gauge theories like QCD.
On-shell methods such as unitarity~\cite{Bern:1994zx,Bern:1994cg}, generalised
unitarity~\cite{Britto:2004nc} and BCFW recursion~\cite{Britto:2005fq} have
been essential tools for deriving the compact five-point two-loop results in
$\mathcal{N}=4$ sYM \cite{Bern:2006vw} and $\mathcal{N}=8$
supergravity~\cite{Carrasco:2011mn}.

Hadron collider phenomenology has gained significant precision thanks to
automated codes that employ numerical reduction procedures to avoid the
traditional bottleneck of large intermediate expressions. The integrand
reduction procedure of Ossola, Papadopoulos and Pittau
(OPP)~\cite{Ossola:2006us} was an important development in this story and has
been used in combination with generalised unitarity or a more traditional
Feynman diagram approach to make phenomenological predictions for a wide range
of complex final states with mass effects.

While extensions of the OPP method to two-loops (or indeed the multi-loop case)
are now known
~\cite{Mastrolia:2011pr,Badger:2012dp,Zhang:2012ce,Mastrolia:2012an,Mastrolia:2012wf,Mastrolia:2013kca}
applications had been previously restricted to specific 'all-plus' helicity
amplitudes~\cite{Badger:2013gxa,Badger:2015lda,Badger:2016ozq} where additional
simplicity led to compact analytic representations. In the general case the
integrand level expressions misses a large number of additional relations
between basis integrals which can be identified from integration-by-parts (IBP)
identities~\cite{Chetyrkin:1981qh}.  The standard method for approaching the
computation of multi-loop amplitudes has been to construct a basis of master
integrals using Laporta's algorithm~\cite{Laporta:2001dd} for solving systems of
IBPs.  This technique quickly becomes complicated for systems with many scales
which has prompted new developments both in optimising current automated
codes~\cite{vonManteuffel:2012np,Smirnov:2014hma,Maierhoefer:2017hyi} and new
approaches suitable for direct use in unitarity based
approaches~\cite{Gluza:2010ws,Larsen:2015ped,Ita:2015tya,Kosower:2018obg}. There
have been successful attempts to include IBP relations directly into the
construction of the amplitudes from unitarity cuts using both the maximal
unitarity method~\cite{Kosower:2011ty,CaronHuot:2012ab} and more recently through numerical
unitarity~\cite{Abreu:2017xsl,Abreu:2017idw}.

Of course the reduction to a basis of integral functions is only part of the problem.
The evaluation of multi-scale integral functions also presents a serious technical challenge.
The first contributions to be completed were the planar topologies with all particles massless
computed using (canonical) differential equation techniques~\cite{Kotikov:1990kg,Gehrmann:1999as,Henn:2013pwa}. Due to bottlenecks in the IBP reduction the non-planar
topologies are still unfinished and are a high priority though many groups are working actively on the problem from different angles~\cite{Chicherin:2017dob,Boehm:2018fpv}.

In the last few months there has been increasing activity on the topic of five particle
scattering amplitudes. The numerical unitarity approach has been used to reduce the planar five-gluon
amplitudes to master integrals using finite field evaluations. Directly solving the system of IBP equations
in the planar case has been possible by optimising the route through the Laporta algorithm to produce
some analytic, though quite lengthy, expressions~\cite{Boels:2018nrr,Chawdhry:2018awn}.

In these proceedings we report on an approach using integrand reduction to obtain analytic formulae for
two-loop five-parton integrands and validate our expressions by providing numerical benchmarks of the integrated amplitudes
for the set independent helicity configurations.

\section{Integrand parametrisations and reconstruction over finite fields}

In a recent paper by one of the authors~\cite{Peraro:2016wsq} it was
demonstrated that numerical sampling of generalised unitarity cuts over finite
fields~\cite{Wang:1981:PAU:800206.806398,Wang:1982:PRR:1089292.1089293,Trager:2006:1145768,RISC3778}
could be used together with integrand reduction to extract analytic
representation of the integrands of complex amplitudes in QCD.

We start by parametrising the partial amplitudes of a standard colour decomposition in terms of irreducible numerators, $\Delta_T$,
\begin{equation}
A^{(2)}\left(1,2,3,4,5\right) = \int
\frac{d^d k_1}{i \pi^{d/2} e^{-\epsilon\gamma_E}} \frac{d^d k_2}{i \pi^{d/2} e^{-\epsilon\gamma_E}}
\, \sum_T
\frac{\Delta_{T}(\{k\},\{p\})}{\prod_{\alpha \in T} D_\alpha},
\label{eq:integrand}
\end{equation}
where $\{k\}=\{k_1,k_2\}$ are the $(d=4-2\eps)$-dimensional loop momenta, $T$
is the set of independent topologies and $\{p\} = \{1,2,3,4,5\}$ are the ordered
external momenta. The index $\alpha$ runs over the set of propagators associated with the topology
$T$. Our planar five-parton amplitudes are built from 57 distinct topologies, giving 425 irreducible numerators when including permutations of the external legs. Each topology, $T$, has an associated irreducible numerator $\Delta_T$ which depends
on Lorentz invariants of the loop momenta and external momenta.

We construct a basis of monomials in these invariants by performing a
transverse decomposition of the loop momentum for each topology along the lines
of the construction of van Neerven and Vermaseren~\cite{vanNeerven:1983vr}.
Specifically we keep track of both 4-dimensional and $(-2\eps)$-dimensional
components in the transverse space: $k_i^\mu = k_{\parallel,i}^\mu +
k^{[4]}_{\perp, i}+k^{[-2\eps]}_{\perp, i}$.  Further details are given in
references ~\cite{Peraro:2016wsq,Badger:2017jhb}. After this decomposition
has been performed the numerator can be parametrised using three classes of
irreducible scalar products (ISPs). ISPs in the parallel space can be written
in terms dot products between loop momenta and external momenta $k_i.p_j$ while
ISPs in the transverse space are either in the 4-d 'spurious' space $k_i.\omega_j$ or extra dimensional space
$\mu_{ij} = -k^{[-2\eps]}_{\perp,i}.k^{[-2\eps]}_{\perp,j}$. We take all external momenta to live in exactly four dimensions.

A basis for $\Delta_T(k_i.p_j,k_i.\omega_j,\mu_{ij})$ is then computed by first finding all possible monomials according
to the gauge theory power counting and then solving a linear system to relate these to an independent basis that does not involve the
$\mu_{ij}$ variables~\cite{Badger:2017jhb}. The choice of ordering within the over-complete set of monomials affects the basis and the
final form of the integrand. This method avoids the polynomial division approach taken in previous
integrand reduction methods~\cite{Zhang:2012ce,Mastrolia:2012an} and the linear system can be
analysed efficiently by sampling over finite fields.

Once the basis of monomials is determined the integrand can be constructed by solving a system of generalised unitarity cuts. On each cut either
a product of on-shell tree-amplitudes or a set of ordered Feynman diagrams can be used as input to an integrand fit using finite field
evaluations. In our first applications of this procedure we have considered both tree-amplitude input using Berends-Giele recursion
relations~\cite{Berends:1987me} in the six-dimensional spinor-helicity approach~\cite{Cheung:2009dc} and Feynman diagrams in which the 't Hooft algebra has been used\footnote{We use QGRAF~\cite{Nogueira:1991ex} to generate Feynman diagrams and FORM~\cite{Kuipers:2012rf,Ruijl:2017dtg}
to perform algebraic manipulations}  to evaluate
the extra-dimensional spinor strings. In the six-dimensional case we perform a dimensional reduction with scalar propagators to
find integrands with including dependence on the spin dimension $g^\mu{}_\mu = d_s$.

In order to control kinematic complexity and to allow for numerical evaluations over finite fields,
we use momentum twistor variables~\cite{Hodges:2009hk} to obtain a rational parametrisation of the multiparticle kinematics.
Analytical representation of our integrand is obtained by performing functional reconstruction from finite fields evaluations.
In this way, we avoid processing large intermediate algebraic expressions that normally appear in the analytical computation.
After the integrand is reconstructed, the transverse space must be integrated~\cite{Mastrolia:2016dhn} to obtain a form
compatible with traditional integration-by-parts (IBP) relations. In this work, we integrate only over spurious space and keep
$\mu_{ij}$ dependence, that can further be removed through dimension shifting identities.

\section{Evaluation of two-loop five-parton amplitudes at benchmark phase-space points}

In this section we provide some benchmark results for five-parton scattering
amplitudes at two loops in QCD. Results are obtained in the leading colour
approximation where only planar diagrams, and integrals, appear.
These results have been obtained using the analytic expressions of the master integrals in~\cite{Gehrmann:2015bfy}, by-passing
the time consuming step of integral evaluations with sector decomposition~\cite{Smirnov:2015mct,Borowka:2017idc} used in our previous
publication.

\subsection{Colour decompositions of five-parton amplitudes}

The amplitudes in this section are defined with the normalisation
\begin{equation}
n= m_\eps N_c \alpha_s/(4\pi),\quad \alpha_s = g_s^2/(4\pi),\quad m_\eps=i (4\pi)^{\eps} e^{-\eps\gamma_E},
\end{equation}
where the dimensional regulator $\eps = \frac{d-4}{2}$, $N_C$ is the number of colours, and $\gamma_E$ is the Euler–Mascheroni constant. Their colour decompositions are given by
\begin{align}
  \mathcal{A}^{(L)}(1_g,2_g,3_g,4_g,5_g) = n^L g_s^3  \sum_{\sigma \in S_5/Z_5} & \tr \left(
  T^{a_{\sigma(1)}} T^{a_{\sigma(2)}} T^{a_{\sigma(3)}} T^{a_{\sigma(4)}} T^{a_{\sigma(5)}} \right) \nonumber\\
& \times
  A^{(L)}\left(\sigma(1)_g,\sigma(2)_g,\sigma(3)_g,\sigma(4)_g,\sigma(5)_g \right)
\end{align}
for five gluons,
\begin{equation}
\mathcal{A}^{(L)}(1_{q},2_{g},3_{g},4_{g},5_{\bar q}) = n^L g_s^3 \sum_{\sigma \in S_3}
 \left( T^{a_{\sigma(2)}} T^{a_{\sigma(3)}} T^{a_{\sigma(4)}} \right)_{i_1}^{\;\;\bar i_5}
 A^{(L)}(1_{q},\sigma(2)_g,\sigma(3)_g,\sigma(4)_g,5_{\bar q})
\end{equation}
for a quark pair and three gluons channel and,
\begin{equation}
\mathcal{A}^{(L)}(1_{q},2_{\bar q},3_{g},4_{Q},5_{\bar Q}) = n^L g_s^3 \bigg[
 \left( T^{a_{3}} \right)_{i_4}^{\;\;\bar i_2} \delta_{i_1}^{\;\;\bar i_5}
 A^{(L)}(1_{q},2_{\bar q},3_g,4_Q,5_{\bar Q})
 + \big( 1 \leftrightarrow 4, 2\leftrightarrow 5 \big)
\bigg]
\end{equation}
for the case of two distinct quark pairs and one gluon. In addition we normalise all amplitudes to the leading order amplitudes
which removes any complex phase,
\begin{equation}
\widehat{A}^{(2)}_{\lambda_1\lambda_2\lambda_3\lambda_4\lambda_5} =%
\frac{A^{(2)}(1^{\lambda_1},2^{\lambda_2},3^{\lambda_3},4^{\lambda_4},5^{\lambda_5})}{%
  A^{(0)}(1^{\lambda_1},2^{\lambda_2},3^{\lambda_3},4^{\lambda_4},5^{\lambda_5})}.
\end{equation}
For helicity configurations that vanish at tree-level the leading term in the expansion around $d=4-2\eps$ and $d_s=4$ at one-loop is used.

\subsection{Evaluation of the master integrals}

The master integrals were computed in~\cite{Gehrmann:2015bfy} using first-order differential equations.
All functions needed are expressed in terms of iterated integrals, where the integration kernels are taken from a set
that was identified in~\cite{Gehrmann:2018yef}. The boundary conditions for the differential equations were
determined by constraints such as the absence of unphysical branch cuts. We determined such boundary points
for each of the physical regions, as well as for the Euclidean region.

Up to weight two, all master integrals are expressed in terms of logarithms and dilogarithms. Weight-three contributions are
expressed in terms of ${\rm Li}_3$ functions and in terms of one-dimensional integrals of logarithms and dilogarithms.
At weight four, we use a representation proposed in~\cite{Caron-Huot:2014lda} that allows to write the functions as a one-fold integral
of known functions, leading to a fast and reliable numerical evaluation, for all kinematic regions.

As a validation of these formulas, we have performed numerical comparisons with~\cite{Papadopoulos:2015jft} and,
for the four-point subtopologies, with~\cite{Gehrmann:2000zt}, finding perfect agreement.

\subsection{Evaluation in the Euclidean region}

We use the phase-space point defined by the invariants
\begin{equation}
s_{12} = -1,\quad
s_{23} = -\frac{37}{78},\quad
s_{34} = -\frac{2023381}{3194997},\quad
s_{45} = -\frac{83}{102},\quad
s_{15} = -\frac{193672}{606645},
\end{equation}
which corresponds to the following values of momentum twistor variables\footnote{The form of the momentum twistor parametrisation is given explicitly
in Reference ~\cite{Badger:2017jhb}},
\begin{equation}
x_{1} = -1,\quad
x_{2} = \frac{79}{90},\quad
x_{3} = \frac{16}{61},\quad
x_{4} = \frac{37}{78},\quad
x_{5} = \frac{83}{102}.
\end{equation}

\begin{table}
\begin{center}
\begin{tabularx}{0.98\textwidth}{|C{1.1}|C{0.5}|C{1.1}|C{1.1}|C{1.1}|C{1.1}|}
  \hline
  &  $\eps^{-4}$ & $\eps^{-3}$ & $\eps^{-2}$ & $\eps^{-1}$ & $\eps^{0}$ \\
  \hline
  $\widehat{A}^{(2),[0]}_{--+++}$ & 12.5 & 27.7526 & -23.7728 & -168.1162 & -175.2103 \\
  \hline
  $\widehat{A}^{(2),[0]}_{-+-++}$ & 12.5 & 27.7526 & 2.5028 & -35.8084 & 69.6695 \\
  \hline
\end{tabularx}
\end{center}
\caption{The (non-zero) leading colour primitive two-loop helicity amplitudes for the $d_s=2$ component of $\widehat{A}^{(2)}(1_g,2_g,3_g,4_g,5_g)$ at the Euclidean
phase space point given in the text.}
\label{tab:ggggg0}
\end{table}

\begin{table}
\begin{center}
\begin{tabularx}{0.98\textwidth}{|C{1.1}|C{0.5}|C{1.1}|C{1.1}|C{1.1}|C{1.1}|}
  \hline
  &  $\eps^{-4}$ & $\eps^{-3}$ & $\eps^{-2}$ & $\eps^{-1}$ & $\eps^{0}$ \\
  \hline
  $\widehat{A}^{(2),[1]}_{+++++}$ & 0 & 0 & -2.5 & -6.4324 & -5.3107 \\
  \hline
  $\widehat{A}^{(2),[1]}_{-++++}$ & 0 & 0 & -2.5 & -12.7492 & -22.0981 \\
  \hline
  $\widehat{A}^{(2),[1]}_{--+++}$ & 0 & -0.625 & -1.8175 & -0.4869 & 3.1270 \\
  \hline
  $\widehat{A}^{(2),[1]}_{-+-++}$ & 0 & -0.625 & -2.7759 & -5.0018 & 0.1807 \\
  \hline
\end{tabularx}
\end{center}
\caption{The leading colour primitive two-loop helicity amplitudes for the $(d_s-2)$ component of $\widehat{A}^{(2)}(1_g,2_g,3_g,4_g,5_g)$ at the Euclidean
phase space point given in the text.}
\label{tab:ggggg1}
\end{table}

\begin{table}
\begin{center}
\begin{tabularx}{0.98\textwidth}{|C{0.08}|C{0.23}|C{0.23}|C{0.23}|C{0.23}|}
  \hline
  & $\widehat{A}^{(2),[2]}_{+++++}$ & $\widehat{A}^{(2),[2]}_{-++++}$ & $\widehat{A}^{(2),[2]}_{--+++}$ & $\widehat{A}^{(2),[2]}_{-+-++}$ \\
  \hline
  $\eps^0$ & 3.6255 &  -0.0664 & 0.2056  & 0.0269 \\
  \hline
\end{tabularx}
\end{center}
\caption{The leading colour primitive two-loop helicity amplitudes for the $(d_s-2)^2$ component of $\widehat{A}^{(2)}(1_g,2_g,3_g,4_g,5_g)$ at the Euclidean
phase space point given in the text.}
\label{tab:ggggg2}
\end{table}

\begin{table}
\begin{center}
\begin{tabularx}{0.98\textwidth}{|C{0.9}|C{0.3}|C{1.2}|C{1.2}|C{1.2}|C{1.2}|}
  \hline
  &  $\eps^{-4}$ & $\eps^{-3}$ & $\eps^{-2}$ & $\eps^{-1}$ & $\eps^{0}$ \\
  \hline
  $\widehat{A}^{(2)}_{++++-}$ & 0 & 0 & -4 & -13.53227 & 6.04865 \\
  \hline
  $\widehat{A}^{(2)}_{+++--}$ & 8 & 7.96829 & -52.39270 & -140.15637 & 47.56872 \\
  \hline
  $\widehat{A}^{(2)}_{++-+-}$ & 8 & 7.96829 & -32.22135 & -47.92349 & 145.97201 \\
  \hline
  $\widehat{A}^{(2)}_{+-++-}$ & 8 & 7.96829 & -40.88511 & -87.02993 & 101.23299  \\
  \hline
\end{tabularx}
\end{center}
\caption{The leading colour primitive two-loop helicity amplitudes for $\widehat{A}^{(2)}(1_q,2_g,3_g,4_g,5_{\bar{q}})$ in the HV scheme at the Euclidean phase space point given in the text.}
\label{tab:qqggg}
\end{table}

\begin{table}
\begin{center}
\begin{tabularx}{0.98\textwidth}{|C{0.9}|C{0.3}|C{1.2}|C{1.2}|C{1.2}|C{1.2}|}
  \hline
  &  $\eps^{-4}$ & $\eps^{-3}$ & $\eps^{-2}$ & $\eps^{-1}$ & $\eps^{0}$ \\
  \hline
  $\widehat{A}^{(2)}_{+-++-}$ & 4.5 & 2.28315 & -32.09848 & -41.39350 & 149.33050 \\
  \hline
  $\widehat{A}^{(2)}_{+--+-}$ & 4.5 & 2.28315 &  -6.32369 & -4.61657  & -32.03278 \\
  \hline
  $\widehat{A}^{(2)}_{+-+-+}$ & 4.5 & 2.28315 & -38.29478 & -43.52329 & 201.04914 \\
  \hline
  $\widehat{A}^{(2)}_{+---+}$ & 4.5 & 2.28315 & -26.71316 & -69.75805 &  22.23653  \\
  \hline
\end{tabularx}
\end{center}
\caption{The leading colour primitive two-loop helicity amplitudes for $\widehat{A}^{(2)}(1_q,2_{\bar{q}},3_g,4_Q,5_{\bar{Q}})$ in the HV scheme at the Euclidean phase space point given in the text.}
\label{tab:qqQQg}
\end{table}

The numerical results are shown in Tables~\ref{tab:ggggg0}~-~\ref{tab:ggggg2},~\ref{tab:qqggg}~and~\ref{tab:qqQQg}  for $ggggg$, $qggg\bar{q}$  and
$q\bar{q}gQ\bar{Q}$ partonic channels, respectively. We have compared the poles of our results against the known universal
IR structure~\cite{Catani:1998bh,Becher:2009qa,Becher:2009cu,Gardi:2009qi}, and the $d_s$ dependence of the IR pole formula in the 5-gluon case is
extracted from the FDH results in~\cite{Gnendiger:2014nxa}.

\subsection{Evaluation in the physical region}

For numerical evaluation in the physical region, we use a phase space point defined by the invariants

\begin{equation}
s_{12} =  \frac{113}{7},\quad
s_{23} = -\frac{152679950}{96934257},\quad
s_{34} =  \frac{1023105842}{138882415},\quad
s_{45} =  \frac{10392723}{3968069},\quad
s_{15} = -\frac{8362}{32585},
\end{equation}
which corresponds to the following values of our momentum twistor variables
\begin{equation}
x_{1} =  \frac{113}{7},\quad
x_{2} = -\frac{2}{9}-\frac{i}{19},\quad
x_{3} = -\frac{1}{7}-\frac{i}{5},\quad
x_{4} =  \frac{1351150}{13847751},\quad
x_{5} = -\frac{91971}{566867}.
\end{equation}

\begin{table}
\begin{center}
\begin{tabularx}{.98\textwidth}{|C{0.7}|C{0.5}|C{1.2}|C{1.2}|C{1.2}|C{1.2}|}
  \hline
  &  $\eps^{-4}$ & $\eps^{-3}$ & $\eps^{-2}$ & $\eps^{-1}$ & $\eps^{0}$ \\
  \hline
  $\widehat{A}^{(2),[0]}_{--+++}$ & 12.5 & -9.17716 + 47.12389 $i$ & -107.40046 - 25.96698 $i$ & 17.24014 - 221.41370 $i$ & 388.44694 - 167.45494 $i$ \\
  \hline
  $\widehat{A}^{(2),[0]}_{-+-++}$ & 12.5 & -9.17716 + 47.12389 $i$ & -111.02853 - 12.85282 $i$ & -39.80016 - 216.36601 $i$ & 342.75366 - 309.25531 $i$ \\
  \hline
\end{tabularx}
\end{center}
\caption{The leading colour primitive two-loop helicity amplitudes for the $d_s=2$ component of $\widehat{A}^{(2)}(1_g,2_g,3_g,4_g,5_g)$ at the physical
phase space point given in the text.}
\label{tab:ggggg0phys}
\end{table}

\begin{table}
\begin{center}
\begin{tabularx}{0.98\textwidth}{|C{0.7}|C{0.5}|C{1.2}|C{1.2}|C{1.2}|C{1.2}|}
  \hline
  &  $\eps^{-4}$ & $\eps^{-3}$ & $\eps^{-2}$ & $\eps^{-1}$ & $\eps^{0}$ \\
  \hline
  $\widehat{A}^{(2),[1]}_{+++++}$ & 0 & 0 & -2.5 & 0.60532 - 12.48936 $i$ & 35.03354 + 9.27449 $i$ \\
  \hline
  $\widehat{A}^{(2),[1]}_{-++++}$ & 0 & 0 & -2.5 & -7.59409 - 2.99885 $i$ & -0.44360 - 20.85875 $i$ \\
  \hline
  $\widehat{A}^{(2),[1]}_{--+++}$ & 0 & -0.625 & -0.65676 - 0.42849 $i$ & -1.02853 + 0.30760 $i$ & -0.55509 - 6.22641 $i$ \\
  \hline
  $\widehat{A}^{(2),[1]}_{-+-++}$ & 0 & -0.625 & -0.45984 - 0.97559 $i$ & 1.44962 + 0.53917 $i$ & -0.62978 + 2.07080 $i$ \\
  \hline
\end{tabularx}
\end{center}
\caption{The leading colour primitive two-loop helicity amplitudes for the $d_s-2$ component of $\widehat{A}^{(2)}(1_g,2_g,3_g,4_g,5_g)$ at the physical
phase space point given in the text.}
\label{tab:ggggg1phys}
\end{table}

\begin{table}
\begin{center}
\begin{tabularx}{0.98\textwidth}{|C{0.1}|C{0.21}|C{0.23}|C{0.23}|C{0.23}|}
  \hline
  & $\widehat{A}^{(2),[2]}_{+++++}$ & $\widehat{A}^{(2),[2]}_{-++++}$ & $\widehat{A}^{(2),[2]}_{--+++}$ & $\widehat{A}^{(2),[2]}_{-+-++}$ \\
  \hline
  $\eps^0$ & 0.60217 - 0.01985 $i$  & -0.10910 - 0.01807 $i$  &  -0.06306 - 0.01305 $i$   & -0.03481 - 0.00699 $i$  \\
  \hline
\end{tabularx}
\end{center}
\caption{The leading colour primitive two-loop helicity amplitudes for the $(d_s-2)^2$ component of $\widehat{A}^{(2)}(1_g,2_g,3_g,4_g,5_g)$ at the physical
phase space point given in the text.}
\label{tab:ggggg2phys}
\end{table}

The numerical results for the $ggggg$ partonic channel are shown in Tables~\ref{tab:ggggg0phys}~-~\ref{tab:ggggg2phys}.

\section{Outlook}

The last few months have seen rapid progress in our ability to compute some of
the missing two-loop amplitudes needed to improve the precision of theoretical
predictions at hadron colliders. While benchmark numerical evaluations have been completed,
the analytic representations of the integrand are extremely large.
Further study of the analytic structure and direct reduction with a complete set of IBPs
will be important to obtain representations suitable for flexible phenomenological applications.
Continuing studies into the structure of amplitudes in maximally supersymmetric gauge theory such
as local integrand structures in planar and non-planar sectors
\cite{Bern:2015ple,Bourjaily:2017wjl,Bern:2017gdk} may prove to give useful insights when persuing this direction.

%While numerical results have been calculated
%the approach produces very large expressions in the analytic case. We see perfect
%agreement with the universal pole structure, however this is not manifest in
%the construction of the integrand parametrisation and there are large
%cancellations between topologies. Understanding how to improve this behaviour may be
%important in order to reduce the complexity and find representations suitable for phenomenological applications.

\section*{Acknowledgements}

This work has received funding from the European Research Council (ERC) under the European Union’s Horizon 2020 research and innovation programme (grant agreements No 725110 and No 772099).
SB is supported by an STFC Rutherford Fellowship ST/L004925/1. BH and CBH are supported by Rutherford Grant ST/M004104/1. We would like the thank the authors of Reference \cite{Abreu:2018jgq} for pointing out an error in the previous version of Table \ref{tab:qqQQg}.

%\bibliographystyle{JHEP-LL12}
%\bibliography{LL2018}

\providecommand{\href}[2]{#2}\begingroup\raggedright\endgroup

\end{document}